\pdfoutput=1
\documentclass[twocolumn,preprintnumbers,superscriptaddress,aps]{revtex4-1}

\usepackage{amsthm,amsmath,amssymb,physics,mathtools}
\usepackage{bbm}

\usepackage{array,esint}
\def\CMD#1{%
   $ \csname#1\endcsname \displaystyle\csname#1\endcsname $ & \texttt{\textbackslash#1} &}

\usepackage{hyperref}
\hypersetup{
    colorlinks = true,       % false: boxed links; true: colored links
    linkcolor = red,          % color of internal links (change box color with linkbordercolor)
    citecolor = magenta,        % color of links to bibliography
    filecolor=red,      % color of file links
    urlcolor= cyan           % color of external links
}

\usepackage{graphicx}
\graphicspath{ {TBA-PRL/} }
\usepackage{xcolor}

\usepackage{dsfont}

\usepackage{mathtools}
\DeclarePairedDelimiter\floor{\lfloor}{\rfloor}

\newcommand{\be}{\begin{equation}}
\newcommand{\ee}{\end{equation}}
\newcommand{\beq}{\begin{equation}}
\newcommand{\eeq}{\end{equation}}
\newcommand{\bea}{\begin{equation} \begin{aligned}}
\newcommand{\eea}{\end{aligned} \end{equation}}

\def\C{\mathcal{C}}

\def\F{\mathcal{F}}

\def\cQ{\mathcal{Q}}

\def\Op{\mathcal{O}}

\def\bP{\mathbb{M}}

\def\R{\mathcal{R}}

\def\u{\chi}
\def\U{U}

\def\Ga{\Gamma_{\alpha}}
\def\Gc{\Gamma_{\textrm{cusp}}}
\def\Go{\Gamma_{\textrm{oct}}}

\def\anset{\boldsymbol{\alpha}_n}
\def\OL{\textrm{1-loop}}
\def\BDS{\textrm{BDS}}

\def\min{\textrm{min}}

\def\adj{\textrm{adj}}

\def\ta{\widehat{a}}
\def\tb{\widehat{b}}
\def\tc{\widehat{c}}

\def\XXint#1#2#3{{\setbox0=\hbox{$#1{#2#3}{\int}$}
     \vcenter{\hbox{$#2#3$}}\kern-.5\wd0}}

\usepackage{comment}
\begin{document}

%\preprint{LPENS-XXXX}

\title{Null limit of large-charge correlators in planar $\mathcal{N}=4$ Super-Yang-Mills theory}
\author{Benjamin Basso}
%\email{benjamin.basso@phys.ens.fr}
\affiliation{Laboratoire de Physique de l'Ecole Normale Sup\'erieure, ENS, Universit\'e PSL, CNRS, Sorbonne Universit\'e, Universit\'e Paris Cit\'e, F-75005 Paris, France}
\author{Thiago Fleury}
%\email{tsi.fleury@gmail.com}
\affiliation{Instituto Cienovus,
Av. Pres. Juscelino Kubitschek, 1327, 4 andar, Conj 41, 
CEP: 04543-011, São Paulo, Brazil}
\author{Erkan Kalu\c{c}}
%\email{osman-erkan.kaluc@ipht.fr}
\affiliation{Laboratoire de Physique de l'Ecole Normale Sup\'erieure, ENS, Universit\'e PSL, CNRS, Sorbonne Universit\'e, Universit\'e Paris Cit\'e, F-75005 Paris, France}
\affiliation{Universit\'e Paris-Saclay, CEA, CNRS, Institut de Physique Th\'eorique
91191 Gif-sur-Yvette, France}
\author{Didina Serban}
%\email{didina.serban@ipht.fr}
\affiliation{Universit\'e Paris-Saclay, CEA, CNRS, Institut de Physique Th\'eorique
91191 Gif-sur-Yvette, France}

\date{\today}

\begin{abstract}
We present a conjecture for the double-logarithmic behavior of the logarithm of large-charge correlators in the null limit in planar $\mathcal{N} = 4$ Super-Yang-Mills theory. Generalizing earlier results for four- and five-point functions, our proposal predicts this behavior for $n$-point functions to all loops in terms of the tilted cusp anomalous dimension. In the dual amplitude description, it reproduces the expected small-mass behavior of massive amplitudes in the equal-mass limit.
\end{abstract}

\maketitle

\section{Introduction}\label{sec:intro}

Correlation functions of chiral primary operators are among the most fundamental observables in planar $\mathcal N=4$ Super-Yang-Mills theory, encoding much of the theory's OPE data. While difficult to determine in general, they exhibit rich structural properties and admit deep connections to scattering amplitudes in special kinematic limits.

\begin{figure}
\begin{center}
\includegraphics[scale=0.26]{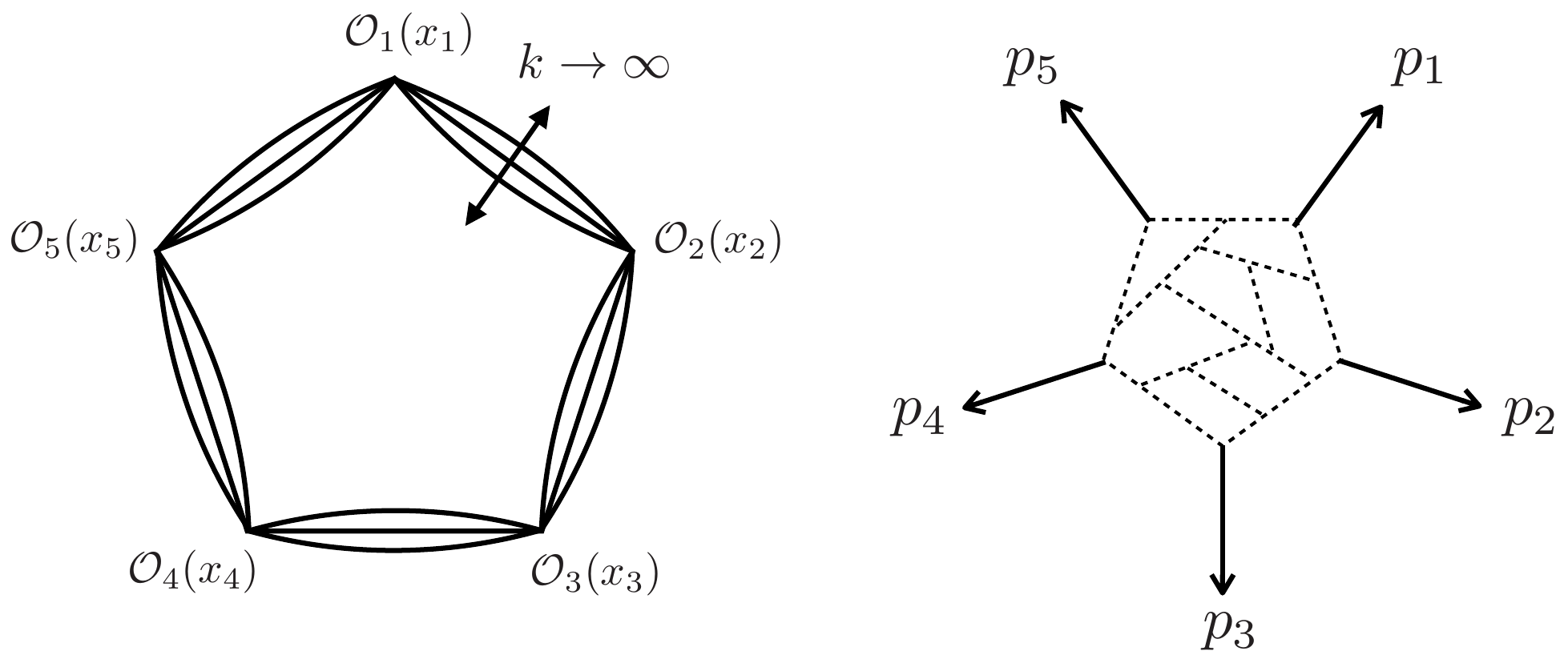}
\end{center}
%\vspace{-2cm}
\caption{Cartoon of the duality between large-charge correlators and scattering amplitudes for $n=5$. The large charge exchanged between operators splits the correlator into two independent polygons in the front and back (left panel). Each polygon is dual to a scattering amplitude in dual momentum space (right panel), with massive external  particles, $p_{i}^2 = x_{i,i+1}^2 \neq 0$, and massless internal ones.}
\label{Polygon} 
\end{figure}

A striking example is provided by the duality between scattering amplitudes and large-charge correlators~\cite{Caron-Huot:2021usw,Bargheer:2025tcw}. It is most naturally formulated by considering the ``ten-dimensional null limit", $(x_{i}-x_{i+1})^2+(y_{i}-y_{i+1})^2\rightarrow 0$, $i = 1, \ldots , n$, of the generating function for correlators of chiral primaries with arbitrary R-charges,
\beq
G_n = \sum_{k_1, \ldots , k_n \geqslant 2} \Big\langle \,\prod_{i\, =\, 1}^{n} \frac{1}{k_i}\textrm{Tr}\, (y_i\cdot \phi(x_i))^{k_i}\, \Big\rangle\, ,
\eeq
where $y\cdot \phi = \sum_{I=1}^{6}y_i^{I}\phi^{I}$ and $y_i^2 = 0$. This limit singles out contributions in which a large R-charge is exchanged between neighboring operators, causing $G_n$ to factorize into the square of a polygon correlator $\bP_{n}(\{x_i, y_i\})$~\cite{Caron-Huot:2021usw,Bargheer:2025tcw}. 
According to a recent proposal, these polygons map in dual momentum space, $p^{\mu}_i = x^{\mu}_{i,i+1}=x^{\mu}_{i}-x^{\mu}_{i+1}$, to Coulomb-branch scattering amplitudes, where the $y_i$ determine the external masses $m_i^2 = (y_i-y_{i+1})^2$, while loop particles remain massless, as illustrated in Fig.~\ref{Polygon}.
This extends to the massive case the well-known factorization of twist-two correlators ($k_i=2$) into Wilson loops, dual to maximally helicity-violating (MHV) amplitudes, in the null-polygon limit~\cite{Alday:2010zy,Alday:2007hr,Drummond:2007aua,Brandhuber:2007yx,Drummond:2007au}.

In the null limit where $m_i^2 = -x_{i,i+1}^2\rightarrow 0$, with all Coulomb-branch parameters set to zero $(y_i\rightarrow 0)$, polygon correlators are expected to behave similarly to massless amplitudes. However, the resulting infrared behavior turns out to be surprisingly different.
Calculations at low multiplicity reveal a double-logarithmic behavior that departs from the Sudakov behavior of conventionally regulated amplitudes~\cite{Bern:2005iz,Alday:2009zm,Henn:2011by,Caron-Huot:2021usw,Alday:2026cpk}, with the infrared logarithms no longer governed by the universal cusp anomalous dimension $\Gc$, but instead by a new set of coefficients characteristic of polygon correlators.

The simplest example of polygon correlator where such a behavior can be observed is the four-point case, $\bP_4 = \mathbb{O}_{0}$, known as the octagon~\cite{Coronado:2018cxj}. Owing to integrability, this quantity can be computed exactly at finite coupling $g = \sqrt{\lambda}/(4\pi)$~\cite{Coronado:2018cxj,Kostov:2019stn,Belitsky:2020qrm}. In the null limit its logarithm exhibits a remarkably simple double-logarithmic scaling 
that persists to all loop orders~\cite{Coronado:2018cxj,Kostov:2019stn,Belitsky:2019fan,Caron-Huot:2021usw},
\begin{align}
\label{eq:exactnulloctagon}
&\log{\bP_4} \\ \nonumber
&\approx -\frac{\Go}{16} \log^2{\left(\prod_{i\, =\, 1}^{4}\frac{x_{i,i+1}^2}{x_{i,i+2}^{2}}\right)} + \frac{g^2}{4} \log^2{\left(\frac{x_{12}^2 x_{34}^2}{x_{23}^2 x_{41}^2}\right)}+\C_4\, ,
\end{align}
up to power-suppressed corrections in the $x_{i, i+1}^2$. Here $\Go$ is the octagon anomalous dimension~\cite{Belitsky:2019fan}
\beq
\Go(g) = \frac{2}{\pi^2} \log{\cosh{(2\pi g)}}\, ,
\eeq
and $\C_4$ is a coupling-dependent constant that is likewise known exactly. Another quantity known in closed form is the one-loop contribution for arbitrary $n$~\cite{Drukker:2008pi,Bargheer:2018jvq},
\beq\label{eq:one-loop-all-n}
\log{\bP_{n}} = g^2\, \bP^{\OL}_{n} +\mathcal{O}\left(g^4\right)\, .
\eeq
In the null-polygon limit, $l_i=\log x_{i,i+1}^2 \rightarrow -\infty$, it exhibits the logarithmic scaling~\cite{Alday:2010zy} (see Appendix~\ref{appx:oneloop})
\beq\label{eq:one-loop-all-n-log}
\!\bP_{n}^{\OL} \approx -\sum_{i\, =\, 1}^{n} (l_i-l'_i)(l_{i+1}-l'_{i}) + 2A_{n}^{\BDS}-\frac{5 n}{2}\zeta_{2}\, ,
\eeq
where $l'_i = \log x_{i,i+2}^2,\,  \zeta_2 = \pi^2/6,$ and $A_{n}^{\BDS}$ coincides with the finite part of the Bern-Dixon-Smirnov (BDS) ansatz for scattering amplitudes~\cite{Bern:2005iz}. Perturbative results for $n=5$, currently available up to three loops~\cite{Bork:2022vat,Bercini:2024pya,Bargheer:2025tcw,Belitsky:2025bgb,Belitsky:2026ukb,Belitsky:2026sci}, further confirmed this behavior and motivated conjectures for the all-loop dependence of the corresponding coefficients~\cite{Belitsky:2026ukb}.

In this Letter, we propose an all-loop formula for the double-logarithmic asymptotics of $\log{\bP_n}$ in the null limit. The formula reproduces all known perturbative results and applies to polygons with an arbitrary number of sides. Its structure parallels that of massless scattering amplitudes in origin limits~\cite{Basso:2020xts,Basso:2022ruw}, and is expressed in terms of the tilted cusp anomalous dimension.

\section{Main formula}\label{sec:mainformula}

Based on the evidence above, we expect that in the null limit $l_{i} = \log{x^2_{i,i+1}}\rightarrow - \infty$, the polygon correlator $\bP_n$ develops large logarithms that exponentiate as
\beq\label{eq:quadratic}
\log{\bP_n} = \cQ_n(\{l_i\}) + \ldots\, ,
\eeq
where $\cQ_n$ is a quadratic form and the dots denotes subleading contributions. This form must respect the discrete dihedral symmetry of the polygon, generated by cyclic shift $l_{i}\rightarrow l_{i+1}$ and reflection $l_{i}\rightarrow l_{n-i+2}$, with $l_{i+n} = l_i$. This forces $\cQ_n$ to be circulant,
\beq
\mathcal{Q}_{n}(\{l_i\}) = \sum_{i, j\,=\,1 }^{n} a_{ij} \,l_i l_j\, , 
\eeq
with $a_{ij} =a(i-j)$ and $a(j) = a(n+j)=a(-j)$. Standard discrete Fourier analysis then diagonalizes it, yielding
\beq\label{eq:quaddivpieceFourier}
\mathcal{Q}_{n}(\{l_i\}) = \frac{1}{n}\sum_{\alpha\, \in\, \anset } \ta_\alpha\,  \Big|\sum_{k\, =\, 1}^{n} e^{2ik\alpha} \, l_k \Big|^2 \, ,
\eeq
where the sum runs over angles $\alpha$ corresponding to the $n$-th roots of unity, $\boldsymbol{\alpha}_n = \{\alpha: e^{2in\alpha} = 1, \alpha\in (-\pi/2, \pi/2] \}$. The Fourier coefficients are defined as
\beq\label{eq:Fourier-transform}
\ta_\alpha \equiv \sum_{j\, =\, 0}^{n-1} a(j)\, e^{2ij\alpha}\, ,
\eeq
and they satisfy $\ta_{\alpha+\pi} = \ta_{\alpha}$ and $\ta_{-\alpha} = \ta_{\alpha}$, reflecting the dihedral symmetry.

Our main conjecture is that the eigenvalue $\ta_\alpha$ is governed by the tilted cusp anomalous dimension $\Ga(g)$,
\beq\label{eq:main}
\ta_\alpha = -\frac{1}{4}  \cos{(2\alpha)} \, \Ga(g) \, ,
\eeq
for arbitrary coupling $g$ and any $n$.

The tilted cusp anomalous dimension $\Ga(g)$ is a function of the tilt angle $\alpha$ and the coupling $g$. It was obtained in~\cite{Basso:2020xts} by deforming the Beisert-Eden-Staudacher (BES) equation for the cusp anomalous dimension~\cite{Beisert:2006ez}, and was originally introduced to describe the double-logarithmic behavior of massless MHV scattering amplitudes in the origin limits, where multiple cross ratios are simultaneously taken to zero. In our setup, the same function controls the null limit of polygon correlators evaluated at discrete angles $\alpha\in \anset$. It obeys $\Ga = \Gamma_{-\alpha} = \Gamma_{\alpha+\pi}$, in agreement with the symmetry properties of $\ta_{\alpha}$.

Explicit expressions for $\Ga(g)$ are obtained by solving the tilted BES equation~\cite{Basso:2020xts}. In particular, at weak coupling one finds
\begin{align}
\label{eq:Ga-weak-coupling}
&\Ga(g) = 4g^2\big\{1-4c^2 \zeta_{2} g^2 +8 c^2(3+5c^2) \zeta_{4} g^4\\
&-8 c^2 \left[(25 + 42c^2 + 35c^4)\zeta_6 + 4 s^2 (\zeta_3)^2\right] g^6 + \mathcal{O}\left(g^8\right) \big\}\, , \nonumber
\end{align}
where $\zeta_{k} = \sum_{j\, =\, 1}^{\infty} j^{-k}$ is the Riemann zeta function and $c = \cos{\alpha}, s = \sin{\alpha}$. Higher-order terms can be generated iteratively.
At special values of $\alpha$, it reproduces known anomalous dimensions, notably $\Gc = \Gamma_{\pm \pi/4}$, $\Go=\Gamma_{0}$, and $\Gamma_{\pi/2} = 4g^2$ to all loops.

Combining this with~\eqref{eq:Fourier-transform} and~\eqref{eq:main}, one immediately verifies consistency with the all-order formula~(\ref{eq:exactnulloctagon}) for $n = 4$. In this case the relevant angles are $\boldsymbol{\alpha}_{4} = \{0, \pm \pi/4, \pi/2\}$, but only $\alpha = 0$ and $\alpha = \pi/2$ contribute to the final result, since the prefactor $\cos{(2\alpha)}$ removes the contribution from $\Gamma_{\textrm{cusp}}$, in complete agreement with~(\ref{eq:exactnulloctagon}).

As an additional check, one reproduces the double-logarithmic behavior $\cQ_n =  -g^2 \sum_{i}l_i l_{i+1}+\ldots$ of the one-loop all-$n$ formula~\eqref{eq:one-loop-all-n-log}. 
At this order, $\Ga$ is independent of $\alpha$, and all angular dependence required to match this behavior comes from the factor $\cos (2\alpha)$ in~\eqref{eq:main}.

\section{Conformal invariance and minimal ansatz}
\label{sectionIII}
Since the original correlator $\bP_{n}$ is a conformally invariant function of the operator positions, it is natural to require that the quadratic form~\eqref{eq:quadratic} be completed into a function of conformal cross ratios.
In principle, for generic $n$, this completion is not unique, and different choices may lead to different subleading divergent terms in~\eqref{eq:quadratic}.
Among these, the simplest and most natural uplift uses only next-to-nearest-neighbor distances $l'_{i} = \log{x^2_{i,i+2}}$.
Accordingly, we consider a decomposition into a divergent and finite part
\beq\label{eq:conf-decomposition}
\log{\bP_{n}} \approx \cQ^{\min}_n + \F_n \, ,
\eeq
where each term is separately conformally invariant. In particular, $\F_n$ is a finite function of the cross ratios of the null polygon.

To construct the ansatz for the divergent part, we generalize $\cQ_n$ to a quadratic form $\mathcal{Q}^{\min}_n$ built form the $2n$ distances $l_i = \log{x^2_{i,i+1}}$ and  $l'_i = \log{x^2_{i,i+2}}$,
\beq\label{eq:conf-ansatz}
\mathcal{Q}^{\min}_{n} = \sum_{i, j}\, a_{ij} l_i l_j + 2\sum_{i, j}\, b_{ij} l_i l'_j + \sum_{i, j} c_{ij}\, l'_i l'_j\, .
\eeq
As before, dihedral symmetry implies that the coefficients depend only on $i-j$ modulo $n$, together with $a(-i) = a(i)$, $b(-i) = b(1+i)$, $c(-i) = c(i)$. To impose conformal invariance, it suffices to consider spacetime inversions, $x_{i}^{\mu} \rightarrow x_{i}^{\mu}/x_{i}^2$, under which
\beq\label{eq:conformal-transformation}
\log{x_{ij}^2}\rightarrow \log{x_{ij}^2} - \log{x_{i}^2} - \log{x_{j}^2}\, .
\eeq
Requiring $\cQ^{\min}_{n}$ to be invariant under this transformation, for arbitrary $\log{x^2_{i}}$ and $\log{x^2_{ij}}$, yields linear recurrence relations among the coefficients in~\eqref{eq:conf-ansatz}, which can be again solved by Fourier transform~\eqref{eq:Fourier-transform}, giving
\beq\label{eq:recurrences}
\begin{aligned}
\tb_\alpha=-e^{i\alpha}\frac{\cos \alpha}{\cos{(2\alpha)}}\,\ta_\alpha\,, \quad \tc_\alpha=\left(\frac{\cos \alpha}{\cos{(2\alpha)}}\right)^2\ta_\alpha\,.
\end{aligned}
\eeq
These expressions are well defined provided $\alpha \neq \pm \pi/4$, or equivalently $n\neq 4k$ with $k\in \mathbb{N}$. Hence, in this case, the conformally invariant quadratic form~\eqref{eq:conf-ansatz} is uniquely determined by the double-divergent part $\cQ_n$.

The result can in fact be directly obtained from~\eqref{eq:quaddivpieceFourier} by replacing $l_i\rightarrow \ell_i$,
\beq\label{eq:tilted-BDS-like}
\cQ^{\textrm{min}}_{n} = -\frac{1}{4n} \sum_{\alpha\, \in\, \anset}\cos{(2\alpha)}\, \Ga \,\Big|\sum_{j\, =\, 1}^{n}e^{2ij\alpha} \,\ell_{j}\Big|^2\, ,
\eeq
where $\ell_i = \log{\u_i}$, with $\u_i$ the unique cross ratio constructed from $x_{i,i+1}^2$ and next-to-nearest-neighbor distances $\{x^2_{j,j+2}\}$~\cite{Alday:2010zy,Alday:2009dv}. It reads
\beq\label{eq:upliftodd}
\u_i = x^2_{i, i+1} \prod_{j\,=\,0}^{2k} \left(x^{2}_{i+2j \sigma, i+2j \sigma+2\sigma}\right)^{\varepsilon(j)}\, ,
\eeq
for $n = 4k+2-\sigma$, with $\sigma=\pm 1$, and
\beq
\!\!\u_i = x^2_{i, i+1} 
\prod_{m\, =\, 0}^1\prod_{j\,=\,0}^{k} \!\left(x^{2}_{i+2j+m, i+2j+m+2}\right)^{\varepsilon(j)/2}\,,
\eeq
for $n=4k+2$, with $k = 1,2,\ldots,$ and $\varepsilon(j) = (-1)^{j+1}$ in both cases.

When $n=4k$, the solution~\eqref{eq:recurrences} is singular due to poles at $\alpha=\pm\pi/4$. These poles do not affect~$\tb_\alpha$, however, because $\ta_\alpha$ has a simple zero at the same points. As a result, the ansatz~\eqref{eq:conf-ansatz} remains well-defined for the full divergent part and arbitrary $n$.

The situation is different for $\tc_\alpha$, where the poles persist. They signal that, for $n=4k$, the divergent part cannot be completed to a conformally invariant quadratic form using only terms of the form $l'_i\, l'_j$. Completion in this case necessarily requires including additional kinematic variables, such as distances between next-to-next-to-nearest neighboring cusps.

This obstruction is tied to the fact that the cross ratios $\u_i$ cannot be defined when $n=4k$~\cite{Alday:2009dv}. It is analogous to the obstruction encountered in the construction of the BDS-like ansatz for massless amplitudes, which formally corresponds to the ansatz~\eqref{eq:main} for $\ta_{\alpha}$ with $\Ga \rightarrow \Gc$. This analogy suggests a natural way to obtain an expression that remains valid for arbitrary $n$, namely by subtracting a contribution controlled by the BDS ansatz and $\Gc$.

In the null limit, this leads to the decomposition
\beq
\label{eq:decompose}
\log{\bP_n} \approx \frac{1}{4}\Gc\,\bP_n^{\OL} + \Delta\cQ^{\min}_n + \R_n\, ,
\eeq
where $\Delta \cQ^{\min}$ is obtained from the ansatz~\eqref{eq:conf-ansatz} by replacing $\Ga$ with $\Delta \Ga = \Ga-\Gc$ in~\eqref{eq:main} and 
solving~\eqref{eq:recurrences} for the modified coefficients. 
The term \(\R_n\) denotes a conformally invariant remainder function, which is finite in the null limit. Unlike $\cQ_n^{\min}$ and $\F_n$, both $\Delta \cQ_n^{\min}$ and $\R_n$ start at two loops in the weak-coupling expansion.

The rationale for this splitting is that the singularities at $\alpha =\pm \pi/4$ are removed from $\Delta \cQ_n^{\min}$, 
the pole in $\tc_\alpha$ being now canceled by the extra zero in $\Delta \Ga$. This allows us to write a conformally invariant formula valid for any $n$ as follows
\beq
\!\!\Delta \cQ^{\min}_{n} = -\frac{1}{16n} \sum_{\alpha\, \in\, \anset} \frac{\Delta\Ga(g)}{\cos{(2\alpha)}}\, \Big|\sum_{j\, =\, 1}^{n}e^{2ij\alpha} \log{\U_j}\Big|^2\, ,
\label{DeltaQmin}
\eeq
with the cross ratios
\beq\label{eq:big-U}
\U_i =  \frac{x^{2}_{i,i+1} x^{2}_{i+2,i+3}}{x^{2}_{i,i+2}x^{2}_{i+1,i+3}}\, .
\eeq
This form is manifestly more local in the index $i$, since the cross ratios $\U_i$ are built only from neighboring points, in contrast to the more non-local cross-ratios $\u_i$. When the latter exist, $\U_i = \u_i\,\u_{i+2}$, and substituting this into~\eqref{DeltaQmin} and~\eqref{eq:decompose} reproduces formula~\eqref{eq:tilted-BDS-like} and~\eqref{eq:conf-decomposition}.

\section{Check in equal-mass limit}

As a further check, we consider the symmetric limit $l_i = l$ for all $i$, with $l\rightarrow -\infty$. In the amplitude picture, this corresponds to the equal-mass regime \(m_i=m\), with \(m\to 0\). In this limit, the correlator should reproduce the factorization formula~\cite{Belitsky:2025bgb,Bork:2022vat,Korchemsky:1988hd,Belitsky:2022itf}
\beq\label{eq:equal-mass}
\log{\bP_n} \approx -\frac{\Go}{4} \sum_{i\, = \, 1}^{n} (l-l'_{i})^2 + F_{n}\, ,
\eeq
where $l = \log{m^2}$ and $F_n$ is another function that remains finite as $m\rightarrow 0$. This expression reflects the factorization of infrared divergences into Sudakov form factors associated with neighboring particles in the planar ordering.
Unlike conventional regulators, however, the divergences are governed by $\Go$ rather than $\Gc$.

This behavior may appear at odds with our proposal, which involves several anomalous dimensions, but in the equal-mass limit $l_j\to l$,
\beq
\sum_{j\, =\, 1}^{n} e^{2ij\alpha} l_{j} \rightarrow n l \delta_{\alpha, 0}\, , \qquad  \alpha \in \anset\, ,
\eeq
where $\delta_{\alpha, 0}$ denotes the Kronecker delta. Consequently, all terms with $\alpha\neq 0$ are projected out, both in the quadratic and linear divergences in $l$ in the ansatz~\eqref{eq:conf-ansatz}. The only surviving contribution is therefore the one proportional to $\Go$, in complete agreement with the structure of~\eqref{eq:equal-mass}.

Equivalently, $\Go$ controls the overall mass dependence, whereas the anomalous dimensions $\Ga$ with $\alpha\neq 0$ encode the dependence on mass ratios, which becomes trivial in the equal-mass limit.

The dependence on $\alpha$ is not entirely lost, however, it survives in the finite part $F_n$. Indeed, our conjecture can be reformulated as a conformal Ward identity for $F_n$. Under the transformation~\eqref{eq:conformal-transformation}, it transforms as 
\beq
F_{n} \rightarrow F_{n} + \sum_{i\, =\, 1}^{n}\delta_i F_{n} \log{x^2_{i}}\, ,
\eeq
to leading order in $\log{x^{2}_{i}}$, with
\beq
\begin{aligned}
\delta_{i} F_{n} &= -\frac{\Go}{2} (l'_i+l'_{i-2}) \\
& \,\,\,\,\,\,+\sum_{\alpha\, \in\, \anset} \frac{\Ga}{n}\cos^2{\alpha} \sum_{j\, =\, 1}^{n} \cos{(2 j\alpha)}\, l'_{j+i-1}\, .
\end{aligned}
\eeq
This equation can be viewed as a deformation of the familiar anomaly equation for the finite part of massless amplitudes~\cite{Drummond:2007au,Alday:2009dv}, which is recovered upon replacing $\Ga, \Go$ with $\Gc$ and summing over the $\alpha$'s.

\section{Weak coupling results}

To perform more advanced checks of our conjecture, it is convenient to first carry out the sum over the angles $\alpha$'s in~\eqref{eq:main} and rewrite the result in the form
\beq\label{eq:unif-conj}
\!\mathcal{Q}_n = -\gamma_{0}(g) \sum_{i\,=\,1}^{n} \, l_{i}l_{i+1} - \sum_{k\, =\, 1}^{\infty} \gamma_{k}(g)\! \sum_{i\,=\,1}^{n} \, l_{i}(l_{i+1+k}+l_{i+1-k})\, ,
\eeq
where $\gamma_{k}$ is the $k$-th Fourier coefficient of $\Ga$, defined as
\beq\label{eq:gammak}
\gamma_{k}(g) = \int\limits_{0}^{\pi/2} \frac{d\alpha}{2\pi} \cos{(2k\alpha)} \, \Ga(g)\, ,
\eeq
or, equivalently,
\beq
\Ga(g) = 4\gamma_{0}(g) + 8\sum_{k\, =\, 1}^{\infty}\cos{(2k\alpha)} \gamma_{k}(g)\, .
\eeq
Note that the dependence on $n$ only enters through the range of the sums in~\eqref{eq:unif-conj} and the periodicity condition $l_i = l_{i+n}$, whereas the coefficients $\gamma_k$ are independent of~$n$.

This representation reveals that, at weak coupling, a new polynomial structure $\propto \bigl(l_i l_{i+L}+l_i l_{i+2-L}\bigr)$ appears at each loop order $L$, coupling logarithms associated with edges separated by up to $L$ steps along the polygon (for $L \leqslant \floor{n/2}$). This follows since the $L$-loop coefficient of $\Ga$ is a polynomial of degree $L-1$ in $\cos{(2\alpha)}$, as can be seen in~\eqref{eq:Ga-weak-coupling}. Consequently, the coefficient~\eqref{eq:gammak} obeys the scaling $\gamma_k=\mathcal{O}(g^{2k+2})$, and at $L$ loops the sum in~\eqref{eq:unif-conj} truncates to $k\leqslant L$.

Up to four loops, the coefficients required are
\begin{align}\label{eq:gamma-coefficients}
&\gamma_{0} = g^2 - 2\zeta_{2} g^4 + 27\zeta_{4} g^6 -\! \left(\frac{627}{2}\zeta_{6} + 4 \zeta_{3}^{2}\right)  g^8 \!+ \!\Op\left(g^{10}\right)\, , \nonumber \\ \nonumber
&\gamma_{1} = -\zeta_2 g^4 + 16 \zeta_4 g^6 - \frac{1597}{8} \zeta_6 g^8 + \Op\left(g^{10}\right)\, , \\ \nonumber
&\gamma_{2} = \frac{5}{2} \zeta_4 g^6 - \left(\frac{189}{4} \zeta_6 - 2\zeta_3^2\right)  g^8 + \Op\left(g^{10}\right)\, , \\
&\gamma_{3} =  - \frac{35}{8} \zeta_6 g^8 + \Op\left(g^{10}\right)\, ,
\end{align}
which fully determine the divergent part for arbitrary $n$.

Using~\eqref{eq:unif-conj}, a perturbative comparison can be made with the available five-point data. These were first obtained up to two loops via bootstrap methods~\cite{Bercini:2024pya,Fleury:2020ykw}, and later confirmed by direct integration of the two-loop five-point integrand in the null limit~\cite{Bargheer:2025tcw,Bargheer:2025uai}; see also~\cite{Kuo:2025bdr} for the corresponding two-loop conformal integrals in full kinematics. These results were further shown to agree with the massless limit of the off-shell five-point amplitude in~\cite{Belitsky:2025bgb,Bork:2022vat}. More recently,~\cite{Belitsky:2026sci,Belitsky:2026ukb} extended the analysis to three loops.

To compare with these results, we restrict~\eqref{eq:unif-conj} to $n=5$ and apply the conformal uplift $l_i \to \ell_i$ as defined in~\eqref{eq:upliftodd}. The conformally invariant quadratic form then takes the general form
\beq\label{eq:decagon}
\!\!\!\cQ^{\min}_5
=c_0 \sum_{i\, =\, 1}^{5}   \ell_{i}^2
+c_1\sum_{i\, =\, 1}^{5} \, \ell_{i}\ell_{i+1}+c_2\sum_{i\, =\, 1}^{5}\ell_{i}\ell_{i+2}\, ,
\eeq
with $\ell_{i+5}= \ell_i$, in agreement with~\cite{Belitsky:2026ukb}, see eqs.~(6.11)--(6.12) therein, upon identifying $\mathbb{L}^2_k (v) = \sum_{i\, =\, 1}^{5} \ell_{i}\ell_{i+k}$. The coefficients $c_k$ follow from matching~\eqref{eq:unif-conj} onto the basis~\eqref{eq:decagon}. Up to four loops, one finds
\beq
\begin{aligned}
&c_0=-\gamma_1(g)\,,\quad c_1=-\gamma_0(g)-\gamma_2(g)-\gamma_3(g)\,,\\
&c_2=-\gamma_1(g)-\gamma_2(g)\,-\gamma_3(g)\,,
\end{aligned}
\eeq
and, using the values in~(\ref{eq:gamma-coefficients}),
\beq
\begin{aligned}
\!\!c_0&=\zeta_2 g^4-16 \zeta_4 g^6+\frac{1597}{8}\zeta_6 g^8\,,\\
\!\!c_1&=-g^2+2 \zeta_2 g^4-\frac{59}{2} \zeta_4 g^6+\left(2\zeta^2_3+\frac{2921}{8}\zeta_6\right)g^8\,,\\ 
\!\!c_2&=\zeta_2 g^4-\frac{37}{2} \zeta_4 g^6+\left(-2 \zeta_3^2 + \frac{1005}{4} \zeta_6\right)g^8\,,
\end{aligned}
\eeq
up to $\mathcal{O}(g^{10})$ corrections. Truncating at three loops, we find perfect agreement with the results of~\cite{Belitsky:2026sci,Belitsky:2026ukb}!

The four-loop expressions are predictions. They differ from those conjectured in~\cite{Belitsky:2026sci,Belitsky:2026ukb} which also rely on the tilted cusp anomalous dimension but involve a different choice of angles $\alpha$. In our construction, the angles are fixed geometrically for any $n$ and, for $n=5$, correspond to the fifth roots of unity, $\boldsymbol{\alpha}_5 = \{0, \pm \pi/5, \pm 2\pi/5\}$.

For completeness, let us also quote the higher-loop conformal formula for generic $n$. As discussed earlier, this is most naturally done using $\Delta \cQ^{\min}$, which starts at two loops and reads
\beq\label{eq:expansion-DeltaQ-secondorder}
\Delta \cQ^{\min}_{n} = \frac{1}{2}\zeta_{2} g^{4}\sum_{i\, =\, 1}^{n} \log^2{\U_i} + \mathcal{O}(g^6)\, ,
\eeq
for any $n$ with $\U_i(=\u_i\u_{i+2})$ defined in~\eqref{eq:big-U}. It would be valuable to check this formula against direct diagrammatic results, using the known two-loop integrand for a generic $n$-gon~\cite{Bargheer:2025tcw,Bargheer:2025uai}. The general all-loop formula reads
\beq\label{eq:expansion-DeltaQ}
\Delta \cQ^{\min}_{n} = -\frac{\tilde{\gamma}_{1}}{2} \sum_{i\, =\, 1}^{n} \log^2{\U_i} - \sum_{k\, = \, 2}^{\infty}\tilde{\gamma}_{k} \sum_{i\, =\, 1}^{n} \log{\U_{i}}\log{\U_{i+k-1}}\, , 
\eeq
with $\tilde{\gamma}_{k} = \sum_{m\, =\, 0}^{\infty}(-1)^{m}\gamma_{k+2m} = \mathcal{O}\left(g^{2k+2}\right)$.

\section{Conclusions}

In this Letter, we proposed an all-loop conjecture for the behavior of polygon correlators in the null limit in planar $\mathcal{N}=4$ Super-Yang-Mills theory. The conjecture rests on a small set of natural assumptions, drawing on the symmetries of the correlator and insights into the general structure of scattering amplitudes in origin limits. It reproduces all available data. Further non-perturbative tests may be obtained using integrability, in particular through the hexagonalization approach to correlation functions~\cite{Inpreparation}.

Uplifting our result to a function of cross ratios through a minimal ansatz, we find that polygon correlators admit the decomposition~\eqref{eq:decompose} in the null limit. This mirrors the twist-two case, where the correlator factorizes into a null polygonal Wilson loop multiplied by a recoil factor that captures the null-limit divergences~\cite{Alday:2010zy,Alday:2013cwa,Bercini:2020msp,Chen:2025ffl}. Our result takes the same form, with
\beq\label{eq:factorization}
J_n  = e^{\Delta \cQ_{n}^{\min}} 
\eeq
playing the role of the recoil factor. Like its twist-two counterpart, $J_n$ depends only on the vanishing cross ratios $\U_i$ defined in~\eqref{eq:big-U} and, at weak coupling, exhibits analogous long-range contributions of the form $\sim g^{2k+2}\log{\U_i}\log{\U_{i+k-1}}$ (see Appendix~\ref{appx:Jlengthtwo}). The large-charge recoil factor is, however, considerably simpler, owing to its Gaussian structure.

An important open question is whether the finite remainder $\R_n$ in~\eqref{eq:decompose} coincides with the remainder function of scattering amplitudes (up to scheme-dependent constants) as for twist-two correlators~\cite{Alday:2010zy,Chen:2025ffl}. We know that the equality with massless amplitudes holds exactly at integrand level~\cite{Caron-Huot:2021usw,Bargheer:2025tcw}. We also expect $\R_n$ (or, more precisely, $\F_n$ in~\eqref{eq:conf-decomposition}) to satisfy the same Steinmann relations and belong to the same space of transcendental functions as MHV amplitudes~\cite{Caron-Huot:2016owq}, suggesting this identification may persist beyond the integrand level.

A key difference, however, is that a large-charge correlator has many particles running along its boundary, rather than the single fast particle that frames the Wilson loop in the twist-two case. This distinction may substantially alter the Wilson-loop picture, leaving a direct interpretation of the remainder less clear.

A direct comparison at two loops is the next natural step, building on the recent results for polygon integrands~\cite{Bargheer:2025tcw,Bargheer:2025uai} 
and off-shell amplitudes~\cite{Belitsky:2025vfc}. The first non-trivial test arises at $n=6$, where a genuine remainder function appears, just as for scattering amplitudes and Wilson loops~\cite{Drummond:2007bm,Bern:2008ap,Drummond:2008aq,Goncharov:2010jf}.

More generally, it would be interesting to study broader classes of Coulomb-branch amplitudes, such as those considered in~\cite{Alday:2025pmg}, which interpolate more smoothly between the off-shell and on-shell regimes. This may help clarify whether the walking behavior identified there is a feature of the divergent part alone or extends to the finite remainder as well.

Finally, another powerful framework for studying the null-polygon limit is the Stampedes approach~\cite{Olivucci:2021pss,Olivucci:2022aza}. It operates in the double-scaling limit, where the cusp variables $t^2_i = g^2 l_{i-1} l_{i}$ are kept finite while $g^2\rightarrow 0$ and $l_i\rightarrow -\infty$, and has enabled nontrivial tests of diagrammatic results at higher loops~\cite{Bercini:2024pya,Belitsky:2025bgb,Bargheer:2025tcw}. While our analysis extends beyond this limit for empty polygons with $y^2_{ij} = (y_i-y_j)^2 \rightarrow 0$, it does not readily apply when some $y^2_{ij}$ remain finite, unlike the Stampedes approach. It would be interesting to explore whether this complementarity can provide access to more general null-polygon configurations at higher loops.

%%%%%%%%%%% Acknowledgements

\begin{acknowledgments}
We thank S.~Caron-Huot, F.~Coronado, L.~Dixon, S.~He, J.~Henn, G.~Korchemsky, and H.~X.~Zhu for inspiring discussions. We acknowledge support from the French National Research Agency through the research grant “Observables” (ANR-24-CE31-7996) and from CNRS through the IRP grant NP-Strong. TF thanks Serrapilheira Institute Grant (Serra–R-2012-38185) for support. DS thanks DESY for hospitality and support.

\end{acknowledgments}

\onecolumngrid
\clearpage

\begin{comment}
% remove 'Supplemental Material' from ToC
\let\oldaddcontentsline\addcontentsline
\renewcommand{\addcontentsline}[3]{}
\section*{Supplemental Material}
\setcounter{subsection}{0}
\let\addcontentsline\oldaddcontentsline

\numberwithin{equation}{subsection}

% footnote
\renewcommand{\thefootnote}{\fnsymbol{footnote}}
\end{comment}

\appendix

\section{One-loop result}
\label{appx:oneloop}

The structure of $n$-point functions of arbitrarily polarized half-BPS operators at one loop was determined in~\cite{Drukker:2008pi} from the complete set of one-loop Feynman diagrams. For polygon correlators, the contributions can be resummed into a closed-form expression, in agreement with both the integrated one-loop polygon integrand of~\cite{Bargheer:2025tcw} and the integrability results of~\cite{Fleury:2017eph,Bargheer:2018jvq} in two-dimensional kinematics. In the limit where all the R-charge contractions $y_{ij}^{2} \rightarrow 0$, the correlator can be expressed as a sum of off-shell box integrals,
\beq
\log \bP^{\OL}_{n} = \frac{1}{2} \sum_{i\, =\, 1}^{n}\sum_{j\, =\, i+2}^{n+i-2} \Phi(z_{ij}, \bar{z}_{ij}) \, ,
\label{P1box}
\eeq
where
\begin{align}
\label{eq:Phidef}
\Phi(z, \bar{z}) = \frac{z+\bar{z}+2}{z-\bar{z}} \left[\textrm{Li}_{2}(-z) +\frac{1}{2}\log{(z\bar{z})}\log{\left(1+z\right)} - (z\leftrightarrow \bar{z})\right]\, ,
\end{align}
and $\textrm{Li}_{2}(x) = \sum_{k\, =\, 1}^{\infty} x^{k}/k^2$ denotes the dilogarithm function. The variables $z_{ij}$ and $\bar{z}_{ij}$ parametrize the cross ratios associated with the four points $x_{i}, x_{i+1}, x_{j}, x_{j+1}$, namely
\beq
U_{ij} = \frac{x^{2}_{i,i+1}x^{2}_{j, j+1}}{x^{2}_{i, j}x^{2}_{i+1, j+1}} = \frac{1}{(1+z_{ij})(1+\bar{z}_{ij})}\, , \qquad V_{ij} = \frac{x^{2}_{i,j+1}x^{2}_{i+1, j}}{x^{2}_{i, j}x^{2}_{i+1, j+1}} = \frac{z_{ij}\bar{z}_{ij}}{(1+z_{ij})(1+\bar{z}_{ij})}\, .
\eeq
Note that this parametrization differs from that commonly used in the literature by a sign flip, $(z,\bar z)\to - (z,\bar z)$. Only $z_{ij}$ and $\bar{z}_{ij}$ with $|i-j|\geq 2$ appear in the sum~\eqref{P1box}.
The null-polygon limit corresponds to $U_{ij}, V_{i,i+2}\to 0$, which implies $z_{ij} \rightarrow \infty$ and $\bar{z}_{i,i+2} \rightarrow 0$, while the remaining $\bar{z}_{ij}$ are held fixed. In this limit,
\beq
\label{eq:limits_cr}
z_{ij} \approx \frac{1-V_{ij}}{U_{ij}} \rightarrow \infty\, , \qquad \bar{z}_{ij} \approx \frac{V_{ij}}{1-V_{ij}}\, .
\eeq
In this appendix, we show that the one-loop contribution $\bP^{\OL}_{n}$ in this limit takes the form~\eqref{eq:one-loop-all-n-log},
\begin{equation} 
\log \bP^{\OL}_{n} \approx
 -\sum_{i\, =\, 1}^n (l_i - l'_i)(l_{i+1} - l'_i) + 2A_n^{\BDS} - \frac{5 n}{2}\zeta_{2}  \, , 
\label{ThePprediction}
\end{equation}
where $l_i = \log{x^2_{i,i+1}}$ and $l^{\prime}_i = \log{x^2_{i,i+2}},$ and $A_{n}^{\BDS}$ denotes the finite part of the BDS ansatz~\cite{Bern:2005iz}. The analysis is essentially the same as that for twist-two correlators~\cite{Alday:2010zy}. We provide here a detailed derivation for completeness.

To translate the results of~\cite{Bern:2005iz} into our notations, we identify $t_i^{[r]} = (k_i + \ldots + k_{i+r-1})^2 \rightarrow x_{i,i+r}^2$.  For $n\geqslant 5$, this yields
\begin{equation}
\label{ABDSn}
 2A^{\BDS}_n =  \sum_{i\, =\, 1}^n g_{n,i} + \frac{3 n}{2}\zeta_{2} \, , 
\end{equation}
where
\begin{equation}
g_{n,i} = - \sum_{r\, =\, 2}^{m -1} \left(l_i^{(r)}-l_i^{(r+1)}\right)
\left(l_{i+1}^{(r)}-l_i^{(r+1)}\right)
 + D_{n,i} + L_{n,i} \, , \qquad m\equiv\floor{n/2}\,,
\end{equation}
with the shorthand notation $l_i^{(r)}\equiv l_{i,i+r}\equiv\log x^2_{i,i+r}$, such that $l_i\equiv l_i^{(1)},\; l'_i\equiv l_i^{(2)}$, and
\begin{align}
\label{eq:DandL}
D_{n,i}&=-\sum_{r\, =\, 2}^{m -1} \textrm{Li}_{2}\,(1-V_{i-1,i+r})+\frac{\delta_{n,2m}}{2}\textrm{Li}_{2}\,(1-V_{i-1,i+m-1})\, , \nonumber \\
L_{n,i}&=-\frac{2-\delta_{n,2m}}{4}\left(l_i^{(m)}-l_{i+m+1}^{(m)}\right)\left(l_{i+1}^{(m)}-l_{i+m}^{(m)}\right)\,.
\end{align}
To avoid lengthy formulas, we included the dependence on the parity of $n$ in some prefactors. Note also that from the definition of the logarithms of distances, we have $l_i^{(r)}=l_{i+r}^{(n-r)}$, so $l_i^{(m)}=l_{i+m}^{(m)}$ for $n$ even, and $l_i^{(m)}=l_{i+m}^{(m+1)}$ for $n$ odd. The case $n=4$ is special and yields
\begin{equation}
\label{ABDS4}
2A^{\BDS}_4 =  \log^2{\left(\frac{x^2_{13}}{x^2_{24}}\right)} + 8\zeta_{2}  \, .  
\end{equation}

There are two relevant asymptotic formulas to consider, namely
\begin{equation}
\begin{aligned}
\Phi(z \rightarrow \infty,\bar{z} \rightarrow 0) \approx - \frac{\pi^2}{6}
+ \frac{1}{2}\log z\, 
\log\bar{z} \, ,  
\end{aligned}
\end{equation}
and
\begin{equation}
\label{eq:Phi_inf}
\begin{aligned}
\Phi(z \rightarrow \infty,\bar{z})  \approx - \frac{\pi^2}{6}
+ \frac{1}{2}\log z\, \log \bar{z}  - \frac{1}{2}
\log (z \bar{z})\, 
\log(1+\bar{z})  - {\rm{Li}}_2\,(- \bar{z}) \, . 
\end{aligned}
\end{equation}
The first case is relevant for $j=i+2$, yielding
\begin{align}
\label{eq:Phiii+2}
\Phi(z_{i,i+2}, \bar z_{i,i+2})\approx -\frac{\pi^2}{6}-\frac{1}{2} \log U_{i,i+2}\,\log V_{i,i+2}\;.
\end{align}
For the second case $|i-j|> 2$, and  $V_{ij} \in [0,1)$ implies $\bar z_{ij}>0$. The argument of the dilogarithm is therefore negative,
and one can use the following two identities, both valid for $x \notin (-\infty, 0]$, 
\begin{equation}
\text{Li}_2\,(-x) + \text{Li}_2\left(-\frac{1}{x}\right) +\frac{\pi^2}{6}= \text{Li}_2\,(1 - x) + \text{Li}_2\left(1 - \frac{1}{x}\right) = - \frac{1}{2} \log^2 x \, ,  %\quad x \notin (-\infty, 0] \, . 
\end{equation}
so that we can write
\begin{equation}
 \text{Li}_2\,(-\bar z_{ij})=\text{Li}_2\,(1- V_{ij}) +R_{ij}\,, 
\qquad R_{ij}\equiv\frac{1}{2}\left(\log^2 V_{ij}-\log^2\left(\frac{1-V_{ij}}{V_{ij}}\right)\right)-\frac{\pi^2}{6}\,.
\end{equation}
The same combination $R_{ij}$ appears in the logarithmic terms in \eqref{eq:Phi_inf}, as can be checked using \eqref{eq:limits_cr}, so that in this case we get
\begin{align}
    \Phi(z_{ij}, \bar z_{ij})\approx -\frac{1}{2} \log U_{ij}\,\log V_{ij}-\text{Li}_2\,(1- V_{ij})\,, \qquad |i-j|> 2\,.
\end{align}
Upon summation and taking into account symmetry factors, the dilogarithmic part reproduces the first line in~\eqref{eq:DandL}. The constant term originates  from~\eqref{eq:Phiii+2} and equals $- n\zeta_2$, for $n>4$, and $- 2\zeta_2$ for $n=4$, in agreement with~\eqref{ABDSn} and~\eqref{ThePprediction}.

The logarithmic terms in the answer are more complicated to repackage. The starting point is
\begin{align}
\label{eq:huge}
    &-\frac{1}{4}\sum_{i\, =\, 1}^{n}\sum_{j\, =\, i+2}^{n+i-2}\log U_{ij}\,\log V_{ij}=
    -\frac{1}{4}\sum_{i\, =\, 1}^{n}\sum_{j\, =\, i+2}^{n+i-2}\left(l_{i,i+1}+l_{j,j+1}-l_{ij}-l_{i+1,j+1}\right)\left(l_{i,j+1}+l_{i+1,j}-l_{ij}-l_{i+1,j+1}\right)
 \nonumber\\
    &=-\frac{1}{4}\sum_{i\, =\, 1}^{n}\sum_{j\, =\, i+2}^{n+i-2}\Big(2l_{i,i+1}(l_{i,j+1}+l_{i+1,j}-l_{ij}-l_{i+1,j+1})+2l_{ij}(l_{ij}+l_{i+1,j+1})-(l_{ij}+l_{i+1,j+1})(l_{i+1,j}+l_{i,j+1})\Big)\,,
\end{align}
where we have used the symmetries $l_{ij}=l_{n+i,j}=l_{ji}$ to reorganize the sums. The first block in the second line of~\eqref{eq:huge} telescopes, leaving
\begin{align}
\label{eq:firstblock}
    -\frac{1}{2}\sum_{i\, =\, 1}^{n}\sum_{j\, =\, i+2}^{n+i-2}l_{i,i+1}(l_{i,j+1}+l_{i+1,j}-l_{ij}-l_{i+1,j+1})&=
    -\frac{1}{2}\sum_{i\, =\, 1}^{n}l_i(l_{i-1}+l_{i+1}-l'_i-l'_{i-1})\\ \nonumber
    &=-{}\sum_{i\, =\, 1}^{n}l_il_{i+1}+\frac{1}{2}\sum_{i\, =\, 1}^{n}l_{i+1}(l'_i+l'_{i+1})\,,
\end{align}
whereas the second block gives, after taking into account symmetries,
\begin{align}
\label{eq:secondblock}
    -\frac{1}{2}&\sum_{i\, =\, 1}^{n}\sum_{j\, =\, i+2}^{n+i-2}l_{ij}(l_{ij}+l_{i+1,j+1})=
    -\sum_{i\, =\, 1}^{n}\sum_{r=2}^{m-1}l_i^{(r)}(l_i^{(r)}+l_{i+1}^{(r)})-\frac{1}{2}(2-\delta_{n,2m})\sum_{i\, =\, 1}^{n}l_i^{(m)}(l_i^{(m)}+l_{i+1}^{(m)})\\ \nonumber = -&\sum_{i\, =\, 1}^{n}\sum_{r=2}^{m-1}l_i^{(r)}l_{i+1}^{(r)}-\sum_{i\, =\, 1}^{n}\sum_{r=1}^{m-2}l_i^{(r+1)}l_{i}^{(r+1)}-\frac{1}{2}(2-\delta_{n,2m})\sum_{i\, =\, 1}^{n}l_i^{(m)}(l_i^{(m)}+l_{i+1}^{(m)})\,.
\end{align}
Lastly, the third block evaluates to
\begin{align}
\label{eq:thirdblock}
    \frac{1}{4}&\sum_{i\, =\, 1}^{n}\sum_{j\, =\, i+2}^{n+i-2}(l_{ij}+l_{i+1,j+1})(l_{i+1,j}+l_{i,j+1})=
    \frac{1}{2}\sum_{i\, =\, 1}^{n}l_{i+1}(l_i'+l'_{i+1})+\frac{1}{2}\sum_{i\, =\, 1}^{n}\sum_{r\, =\, 2}^{n-3}l_{i+1}^{(r)}(l_i^{(r+1)}+l_{i+1}^{(r+1)})\\
     =\frac{1}{2}&\sum_{i\, =\, 1}^{n}l_{i+1}(l_i'+l'_{i+1})+\sum_{i\, =\, 1}^{n}\sum_{r\, =\, 2}^{m-1}l_{i+1}^{(r)}(l_i^{(r+1)}+l_{i+1}^{(r+1)})+\frac{1}{2}(1-\delta_{n, 2m})
    \sum_{i\, =\, 1}^{n}l_{i+1}^{(m)}(l_i^{(m+1)}+l_{i+1}^{(m+1)})\,, \nonumber
\end{align}
where in the first equality we used $l_{i+1,j}=l_{j,i+1}$ and exchanged summation indices $i\leftrightarrow j$, and in the second one we used the symmetry of the second sum under $r\leftrightarrow n-r-1$ to combine various terms together. To highlight the divergent terms, we employed the notations $l_i$ and $l'_i$ for them. We compare the result from putting together the last lines of equations~\eqref{eq:firstblock}, \eqref{eq:secondblock} and~\eqref{eq:thirdblock} with the corresponding content of the ansatz~\eqref{ThePprediction},
\begin{equation}
\label{eq:divandg}
- \sum_{i\, =\, 1}^{n}\sum_{r\, =\, 1}^{m -1} \left(l_i^{(r)}-l_i^{(r+1)}\right)
\left(l_{i+1}^{(r)}-l_i^{(r+1)}\right)
=- \sum_{i\, =\, 1}^{n}\sum_{r\, =\, 1}^{m -1}\left(l_i^{(r)}l_{i+1}^{(r)}+l_i^{(r+1)}l_i^{(r+1)}- l_i^{(r+1)}(l_{i}^{(r)}+l_{i+1}^{(r)})\right)\,.
\end{equation}
The mismatch between the two expressions~\eqref{eq:huge} and~\eqref{eq:divandg} is captured by $L_{n,i}$, as defined in~\eqref{eq:DandL},
\begin{align}
\sum_{i\, =\, 1}^{n}l_i^{(m)}l_i^{(m)}-\frac{1}{2}\sum_{i\, =\, 1}^{n}l_i^{(m)}(l_i^{(m)}+l_{i+1}^{(m)})=-\frac{1}{4}\sum_{i\, =\, 1}^{n}\left(l_i^{(m)}-l_{i+1}^{(m)}\right)\left(l_{i+1}^{(m)}-l_{i}^{(m)}\right)= \sum_{i\, =\, 1}^{n}L_{n,i}\,,
\end{align}
for $n=2m$, and 
\begin{align}
-\sum_{i\, =\, 1}^{n}l_i^{(m)}l_{i+1}^{(m)}+\frac{1}{2}\sum_{i\, =\, 1}^{n}l_{i+1}^{(m)}(l_i^{(m+1)}+l_{i+1}^{(m+1)})=-\frac{1}{2}\sum_{i\, =\, 1}^{n}\left(l_i^{(m)}-l_{i}^{(m+1)}\right)\left(l_{i+1}^{(m)}-l_{i-1}^{(m+1)}\right)= \sum_{i\, =\, 1}^{n}L_{n,i}\,,
\end{align}
for $n=2m+1$. For odd $n$, we used $l_i^{(m)}=l_{i+m}^{(m+1)}$ in the second equality, and $\sum_il_{i+1}^{(m+1)}\,l_{i+1}^{(m)}=\sum_il_{i+1}^{(m)}\,l_{i}^{(m+1)}$ in the first.

\section{Comparison with $J$ factor of twist-two operators}
\label{appx:Jlengthtwo}

In this appendix, we compare our results with the known expression for correlators of twist-two operators in the null-polygon limit. General arguments imply that, in this limit, the correlator factorizes into a null polygonal Wilson loop in the adjoint representation multiplied by a recoil factor $J^{\adj}_{n}$~\cite{Alday:2010zy,Alday:2013cwa,Bercini:2020msp,Chen:2025ffl}. This factor captures the interaction between the state propagating along the Wilson-loop contour and the color flux tube it generates. It was first computed at two loops in~\cite{Alday:2010zy}. Its all-loop expression was derived in~\cite{Alday:2013cwa} for $n=4$ using conformal-bootstrap techniques, and generalized to higher $n$-point functions in~\cite{Bercini:2020msp}.

In terms of the cross-ratios $\U_i$ defined in~\eqref{eq:big-U}, the recoil factor takes the form
\beq\label{eq:J-twist-two}
J^{\adj}_{n} = e^{\sum_{i=1}^{n}\Delta_{i,i+1}}\prod_{i\, =\, 1}^{n} G\left(\frac{f}{4}\log{\U_i} -\frac{h}{2}\right)\, ,
\eeq
where $\Delta_{i, i+1}$ is a diffusion operator acting on the variables $\log{U_i}$,
\beq
\Delta_{i,i+1} = -\frac{4}{f}\frac{\partial^2}{\partial \log{\U_i}\,\partial \log{\U_{i+1}}}\, ,
\eeq
and
\beq\label{eq:gseries}
G(x) = e^{\gamma_{\textrm{E}}x}\Gamma(1+x) = \exp\left[\sum_{k\, =\, 2}^{\infty}\frac{(-x)^k}{k}\zeta_k\right]\, .
\eeq
Here, $\Gamma$ denotes the Gamma function and $\gamma_{\textrm{E}}$ the Euler--Mascheroni constant. Moreover, $f = \Gc^{\adj} = 2\Gc$ is the cusp anomalous dimension in the adjoint representation, while $h = \mathcal{O}\left(g^{4}\right)$ denotes the virtual anomalous dimension of twist-two operators, which is known to all loop orders from integrability~\cite{Freyhult:2009my,Fioravanti:2009xt}.

The logarithm of this factor, $\log J^{\textrm{adj}}_{n}$, plays a role analogous to our conformal ansatz $\Delta \mathcal{Q}_n^{\min}$. Both quantities depend on the cross ratios $U_i$ defined in~\eqref{eq:big-U}, which vanish in the null limit and parametrize the approach to the null polygon. Since the operators inserted at the cusps are different in the two setups, there is no reason to expect the corresponding expressions to coincide. In particular, $\log J^{\textrm{adj}}_{n}$ is not a quadratic function of logarithms at higher loops. Nevertheless, at weak coupling the two quantities exhibit remarkably similar long-range contributions of the form $\log \U_i\,\log \U_{i+k-1}$. These terms appear at order $\mathcal{O}\left(g^{2k+2}\right)$ in both cases and come with closely related coefficients.

To see this, we may set $h\to 0$ in~\eqref{eq:J-twist-two}, since $h$ affects only subleading logarithmic terms at a given loop order. We may also truncate the expansion~\eqref{eq:gseries} at quadratic order,
\beq
G(x) \longrightarrow \exp\left[\frac{x^2}{2}\zeta_2\right]\, ,
\eeq
since higher powers of $x$ contribute to the long-range quadratic terms only through subleading powers of the coupling. The term proportional to $\log{\U_i}\log{\U_{i+k-1}}$ in $\log{J^{\adj}_{n}}$ then arise from the action of the chain of diffusion operators
\beq
\Delta_{i,i+1}
\Delta_{i+1,i+2}
\cdots
\Delta_{i+k-2,i+k-1}
\eeq
on the monomials
\beq
\left(
\frac{f^2}{32}\zeta_2
\right)^{k}
\log^2{\U_i}
\log^2{\U_{i+1}}
\cdots
\log^2{\U_{i+k-1}}
\subset
\prod_{i\, =\, 1}^{n}G\left(\frac{f}{4}\log{\U_i} -\frac{h}{2}\right)\, .
\eeq
Hence, focusing on the quadratic terms, we may write
\beq
\begin{aligned}\label{eq:quadraticJadj}
\log{J^{\adj}_{n}}\big|^{\textrm{Gaussian}} = -\frac{1}{2} d_{1} \sum_{i\, =\, 1}^{n} \log^{2}{\U_i} - \sum_{k\, =\, 2}^{\infty} d_k \sum_{i\, =\, 1}^{n} \log{\U_i}\log{\U_{i+k-1}}\, ,
\end{aligned}
\eeq
with
\beq\label{eq:d-coefficients}
d_{k} \approx (-1)^{k} \frac{f^{k+1}}{4^{k+1}} \zeta_2^{k}\, ,
\eeq
to leading order at weak coupling, $f = \mathcal{O}\left(g^{2}\right)$. This is precisely the structure observed in~\eqref{eq:expansion-DeltaQ-secondorder} for $\Delta Q_{n}^{\min}$. The fact that the same pattern emerges in two rather different setups suggests that the scaling of long-range contributions with the coupling may be a universal feature of the null limit of correlators of local operators. It would be interesting to understand whether this behavior admits a simple diagrammatic interpretation or a reformulation within the framework of~\cite{Alday:2010zy,Chen:2025ffl}.

Finally, it is worth noting that the coefficients in~\eqref{eq:d-coefficients} and~\eqref{eq:expansion-DeltaQ-secondorder} match \textit{exactly} at weak coupling upon formally replacing
\beq\label{eq:rep-f}
f\rightarrow \Gc \approx 4g^2\, ,
\eeq
that is, by using the cusp anomalous dimension appropriate to a fundamental Wilson loop rather than to a Wilson loop in the adjoint representation. Indeed, under this replacement one finds $d_{k}\approx \tilde{\gamma}_k$ with
\beq\label{eq:gammakleadingorder}
\tilde{\gamma}_{k} \approx \gamma_{k} \approx (-1)^{k}\zeta_2^{k}g^{2k+2}\, ,
\eeq
to leading order at weak coupling. This result follows from the definition of $\gamma_{k}$ as the Fourier coefficient of $\Ga$ (see eq.~\eqref{eq:gammak}) and from the fact that the $L$-loop contribution in $\Ga(g) = \sum_{L=1}^{\infty}\Gamma^{(L)}_{\alpha} g^{2L}$ has the form
\beq
\Gamma^{(L)}_{\alpha} = 4(-2\zeta_{2}\cos^2{\alpha})^{L-1} + \ldots\, ,
\eeq
where the dots denote terms with lower powers of $\cos^2{\alpha}$.

\bibliography{NullPolygonsl}
\end{document}